\documentclass[letterpaper,conference]{IEEEtran}
\usepackage{cite}
\usepackage{amsmath,amssymb,amsfonts}
\usepackage{algorithmic}
\usepackage{graphicx}
\usepackage{textcomp}
\usepackage[table]{xcolor}
\usepackage[caption=false]{subfig}
\usepackage{amsmath}
\usepackage{multirow}
\usepackage{soul}
\usepackage[]{nohyperref}  
\usepackage{url}  

\def\BibTeX{{\rm B\kern-.05em{\sc i\kern-.025em b}\kern-.08em
	T\kern-.1667em\lower.7ex\hbox{E}\kern-.125emX}}
\begin{document}
\title{Integrating LoRaWAN with Mobile Ad-hoc Networks for Enhanced Campus Communication}
\author{Ramakant Kumar}
                \maketitle
\begin{abstract}
The integration of Long Range Wide Area Network (LoRaWAN) with Mobile Ad-hoc Networks (MANETs) presents a promising solution for enhancing communication networks within campus environments. This paper explores the unique advantages of combining these two technologies, including scalability, energy efficiency, flexibility, and support for diverse applications. LoRaWAN’s low power consumption and extended range capabilities address the challenges of traditional communication methods, enabling reliable data transmission across various campus scenarios, such as emergency alerts, event coordination, and real-time monitoring. We also identify key challenges faced in this integrated architecture, including signal interference, data packet collisions, and energy management. By providing a comprehensive survey of existing techniques and solutions categorized by the network protocol stack layers, this study aims to inform future research and development efforts in creating robust, energy-efficient communication systems tailored for modern educational institutions. Ultimately, the findings highlight the potential of LoRaWAN-MANET architectures to transform campus communication into a more reliable, adaptable, and cost-effective framework.
\end{abstract}
\section{Introduction}
Mobile Ad-hoc Networks (MANETs) are self-organizing, self-configuring, infrastructure-less wireless networks that allow nodes to communicate directly or through intermediate nodes, without depending on a centralized infrastructure. When combined with LoRaWAN (Long Range Wide Area Network) technology, MANETs become even more powerful, supporting long-range, low-power communication networks ideal for large-scale and dynamic environments like institute campuses~\cite{9130098,8502812,8326735,conf/wcnc/ThangadoraiSGK24,conf/networking/PandeyKGRR24,conf/wcnc/PandeyKG024}.

In a campus setting, LoRaWAN-enabled MANETs can enable secure and efficient real-time communication for a variety of applications, including emergency alerts, event management, and environmental monitoring. For instance, in an emergency or during events, each device (such as smartphones, sensors, or portable communication units) can dynamically connect to nearby nodes to relay messages across the campus. With LoRaWAN’s range of several kilometers, nodes can reduce intermediate hops—critical in traditional MANETs that often rely on Wi-Fi's shorter range.

The infrastructure-free nature of LoRaWAN-MANETs also makes them ideal for challenging environments like disaster response, military operations, and mining. However, despite their advantages, these networks face challenges such as limited energy resources, a dynamic topology from node mobility, low bandwidth, and security concerns. Energy efficiency, in particular, is critical since LoRaWAN nodes often operate on batteries. 

This paper provides a comprehensive survey of existing techniques for addressing these limitations, categorizing them based on different layers of the network protocol stack, including energy-efficient routing, dynamic topology management, secure protocols, and methods for optimizing data transmission in low-power environments~\cite{conf/wcnc/0001G24,10122600,10176016,journals/tpds/MishraGBD24,conf/sensys/KumariG023}. 

The integration of LoRa technology with MANETs combines the benefits of self-organizing, infrastructure-less networks with LoRa’s extended range and low-power attributes, offering a robust, energy-efficient solution for large campuses and remote areas where scalability, power conservation, and adaptability are essential.

\section{Layers Architecture of LoRaWAN-MANETs}
The layered architecture of LoRaWAN-MANETs combines the MANET structure with LoRaWAN’s protocol stack, ensuring efficient data handling across layers and supporting the unique demands of an infrastructure-less, long-range, low-power communication network~\cite{gupta2024lessonslearnedsmartcampus}. This architecture can be organized into the following layers, each responsible for handling different functions essential for reliable and energy-efficient operations in dynamic environments like campuses or remote sites~\cite{10122600,10176016,1321026}.

\subsection{Physical Layer of LoRaWAN-MANETs}

The Physical Layer in LoRaWAN-MANETs is the foundation for data transmission, responsible for managing signal modulation, power control, and channel access across long-range, low-power communication links. By using LoRa modulation techniques such as Chirp Spread Spectrum (CSS), this layer enables LoRaWAN-MANETs to support reliable, energy-efficient communication over several kilometers, making it particularly suitable for applications within large-scale environments, like institute campuses.

\subsubsection{Key Features of the Physical Layer}

\begin{itemize}
    \item \textbf{Long-Range Communication with Chirp Spread Spectrum (CSS):}
    \begin{itemize}
        \item \textbf{Chirp Spread Spectrum Modulation:} LoRaWAN utilizes CSS, a modulation technique that spreads the data across a wider bandwidth, enhancing resilience to noise and interference. This enables nodes to transmit data over distances up to 10-15 km in rural settings and around 2-5 km in urban or campus environments~\cite{1589116,4460126,5558084,5567086}.
        \item \textbf{Adaptive Transmission:} The physical layer can adjust the transmission range and data rate based on the environment, allowing devices to communicate effectively even when nodes are dispersed across a large campus.
    \end{itemize}

    \item \textbf{Low Power Consumption:}
    \begin{itemize}
        \item \textbf{Energy Efficiency:} The LoRa Physical Layer is designed to be extremely energy-efficient, requiring minimal power for data transmission. Nodes can often operate for years on a single battery, making them ideal for deployment in a MANET where continuous recharging may be impractical.
        \item \textbf{Duty Cycling:} To conserve energy, the physical layer supports low-duty cycle operation, allowing nodes to remain in a sleep state when not actively transmitting, reducing power consumption significantly.
    \end{itemize}

    \item \textbf{Data Rate and Bandwidth Adaptation:}
    \begin{itemize}
        \item \textbf{Adaptive Data Rate (ADR):} LoRaWAN’s ADR mechanism automatically adjusts data rates based on the quality of the link, ensuring optimal performance while balancing range, data throughput, and energy use. This adaptability is crucial on a large campus where environmental and physical obstacles, such as buildings, affect signal propagation~\cite{6054047,6780609,7123563,7377400}.
        \item \textbf{Bandwidth Flexibility:} The physical layer operates on several ISM frequency bands (e.g., 868 MHz in Europe, 915 MHz in North America), allowing compatibility across different geographic locations. This flexibility ensures that LoRaWAN-MANETs can adapt to different regulatory environments while maximizing network performance.
    \end{itemize}

    \item \textbf{Interference Resistance:}
    \begin{itemize}
        \item \textbf{Robust Signal Processing:} The spread-spectrum nature of CSS provides high interference resistance, allowing the network to coexist with other wireless devices operating on the same unlicensed band~\cite{7460727,7488250,7498684,7745306}. This is beneficial in campus settings where Wi-Fi, Bluetooth, and other RF devices are prevalent.
    \end{itemize}
\end{itemize}

\subsubsection{Campus Scenario: LoRaWAN-MANETs for Campus-Wide Communication}

In an institute campus scenario, the Physical Layer of LoRaWAN-MANETs plays a vital role in ensuring seamless, long-range communication across diverse applications. Consider the following example applications within a campus:

\begin{itemize}
    \item \textbf{Emergency Alerts and Safety Notifications:} During an emergency, such as a fire or security incident, LoRaWAN-enabled devices across campus (e.g., in student dorms, lecture halls, and labs) can transmit alerts over long distances, ensuring that notifications reach all areas of the campus quickly. With CSS modulation, these alerts can travel across the campus regardless of interference from other networks or the physical obstacles posed by buildings~\cite{7803607,7815384,7880946}.

    \item \textbf{Environmental Monitoring and Smart Infrastructure:} LoRaWAN-MANETs can be deployed for environmental monitoring, tracking parameters like air quality, noise levels, and temperature across campus. For instance, sensors placed in different campus zones communicate their readings through the LoRa Physical Layer, transmitting data over long distances without needing intermediate relay nodes~\cite{conf/icc/KumariGD20,conf/icc/KumariGDB23,conf/iccps/0012G023}. This long-range capability reduces the need for densely placed infrastructure, minimizing maintenance and power requirements.

    \item \textbf{Event Management and Real-Time Coordination:} During large campus events, such as sports meets or convocations, the Physical Layer’s low-power operation enables temporary, ad-hoc networks that connect various nodes (e.g., event staff, security personnel, or information booths). LoRaWAN’s long-range communication allows staff to remain connected across wide event areas with minimal power usage, facilitating real-time updates and coordination.
\end{itemize}

\subsubsection{Challenges and Solutions in the Physical Layer}

Despite its advantages, the Physical Layer of LoRaWAN-MANETs in a campus scenario faces challenges:

\begin{itemize}
    \item \textbf{Obstacle Interference:} Campus buildings and natural features like trees can obstruct LoRa signals. To mitigate this, the network can optimize node placement, for example, by positioning gateway nodes on rooftops or in open areas to maintain line-of-sight communication~\cite{conf/infocom/KumariGD21,conf/infocom/MishraGD22,conf/mswim/KumariGM021,conf/networking/PandeyKGRR24}. The Adaptive Data Rate (ADR) mechanism can further adjust the data rate to accommodate different propagation conditions.

    \item \textbf{Coexistence with Other Wireless Networks:} With a campus saturated with Wi-Fi, Bluetooth, and cellular networks, managing interference is critical. The Physical Layer’s CSS modulation provides resilience against interference, allowing LoRa to coexist with other networks. Additionally, selective channel allocation within the ISM band can help reduce potential conflicts with other wireless devices.

    \item \textbf{Power Management in Battery-Dependent Devices:} Nodes across campus may be deployed in locations where regular battery replacement is challenging~\cite{conf/sensys/KumariG023,conf/wcnc/0001G24,conf/wcnc/ThangadoraiSGK24,conf/wcnc/PandeyKG024}. The duty cycling and low-power operation of the Physical Layer minimize energy use, extending the operational life of battery-powered nodes. For high-use areas, nodes can be solar-powered or connected to low-voltage campus infrastructure.
\end{itemize}

\subsection{Data Link Layer of LoRaWAN-MANETs}

The Data Link Layer in LoRaWAN-MANETs is responsible for ensuring reliable data transfer across nodes, managing error correction, packet formatting, and Medium Access Control (MAC). In a campus setting, the Data Link Layer plays a crucial role in controlling access to the shared wireless medium, especially where multiple devices need to communicate over limited bandwidth.

\subsubsection{Key Features of the Data Link Layer}

\begin{itemize}
    \item \textbf{Error Detection and Correction:}
    \begin{itemize}
        \item LoRaWAN employs cyclic redundancy checks (CRC) at this layer to detect errors in transmitted data packets, ensuring the integrity of messages exchanged across campus.
    \end{itemize}

    \item \textbf{Adaptive MAC Protocol:}
    \begin{itemize}
        \item LoRaWAN employs an Aloha-based MAC protocol, allowing devices to transmit without synchronization. While this is simple and effective, it can lead to packet collisions in a busy campus environment.
        \item \textbf{Duty Cycle Management:} To avoid network congestion, the duty cycle limit regulates the time each device can transmit, balancing load and reducing packet collisions across the network.
    \end{itemize}
\end{itemize}

\subsubsection{Campus Scenario for the Data Link Layer}

In a campus scenario, the Data Link Layer enables seamless communication by managing collisions and ensuring data integrity. For example, in a large event setting, such as a seminar, where multiple nodes attempt to send updates simultaneously, the MAC protocol ensures efficient medium access to avoid packet loss.

\subsubsection{Challenges and Solutions}

\begin{itemize}
    \item \textbf{Collision Management:} Given the high density of devices in a campus, packet collisions may occur frequently. Solutions include scheduling mechanisms or implementing dynamic backoff strategies to minimize congestion.
    \item \textbf{Scalability:} The Aloha MAC protocol may struggle with high device counts. Adaptive access control and scheduling algorithms can help improve scalability for campus-wide deployment.
\end{itemize}

\subsection{Network Layer of LoRaWAN-MANETs}

The Network Layer in LoRaWAN-MANETs is responsible for routing packets across nodes in a multi-hop network, handling address allocation, and managing mobility, particularly vital for a dynamic campus environment.

\subsubsection{Key Features of the Network Layer}

\begin{itemize}
    \item \textbf{Routing Protocols:} 
    \begin{itemize}
        \item MANETs in a campus scenario often use on-demand routing protocols such as AODV (Ad hoc On-Demand Distance Vector), which establish routes as needed to conserve energy and bandwidth.
    \end{itemize}

    \item \textbf{Mobility Management:} 
    \begin{itemize}
        \item The Network Layer adjusts routes dynamically as students and staff move around campus, ensuring data packets are relayed through available nodes in real time.
    \end{itemize}
\end{itemize}

\subsubsection{Campus Scenario for the Network Layer}

In a campus, the Network Layer enables seamless mobility support as people move across buildings and open spaces. For instance, during a fire drill, the network must efficiently redirect communication as individuals move, maintaining communication despite rapid topology changes.

\subsubsection{Challenges and Solutions}

\begin{itemize}
    \item \textbf{Dynamic Topology:} Frequent node movement can disrupt established routes. On-demand routing protocols, combined with periodic route updates, help maintain stability in this dynamic environment.
    \item \textbf{Energy-Efficient Routing:} Routing decisions are made with energy conservation in mind, choosing the shortest or least energy-consuming path to prolong node battery life.
\end{itemize}

\subsection{Transport Layer of LoRaWAN-MANETs}

The Transport Layer is essential for managing end-to-end data transfer, ensuring reliability, flow control, and error correction in the MANET environment. This layer is especially critical for applications that require guaranteed data delivery~\cite{conf/wowmom/KumariGDD22,conf/wowmom/MishraKGSDSP21,journals/cem/GhoshMGSR22}.

\subsubsection{Key Features of the Transport Layer}

\begin{itemize}
    \item \textbf{Reliable Data Transfer:}
    \begin{itemize}
        \item LoRaWAN can use protocols like UDP for applications that prioritize low-latency over reliability and lightweight reliability mechanisms for critical data, balancing efficiency and reliability.
    \end{itemize}

    \item \textbf{Flow Control and Congestion Management:}
    \begin{itemize}
        \item The Transport Layer manages the rate of data flow to avoid congestion on the network, a valuable feature in high-density campus settings where numerous devices may be transmitting.
    \end{itemize}
\end{itemize}

\subsubsection{Campus Scenario for the Transport Layer}

In a campus environment, the Transport Layer can ensure that high-priority alerts, such as security messages, are reliably delivered without congestion issues. For example, in a lab experiment monitoring system, consistent data transmission is critical to prevent data loss or delay.

\subsubsection{Challenges and Solutions}

\begin{itemize}
    \item \textbf{Congestion Management:} High-density networks require effective congestion control to avoid data packet loss. The use of flow control protocols ensures data integrity.
    \item \textbf{End-to-End Reliability:} Applications needing reliable data transfer can use selective acknowledgments and retransmission mechanisms.
\end{itemize}

\subsection{Application Layer of LoRaWAN-MANETs}

The Application Layer in LoRaWAN-MANETs is where various applications, such as remote monitoring, alert systems, and environmental sensors, interface with the network.

\subsubsection{Key Features of the Application Layer}

\begin{itemize}
    \item \textbf{Support for Multiple Applications:} 
    \begin{itemize}
        \item Applications tailored for campus settings, such as emergency alert systems, asset tracking, and environmental monitoring, can be supported by the Application Layer.
    \end{itemize}
    
    \item \textbf{Data Aggregation and Compression:} 
    \begin{itemize}
        \item The Application Layer supports data aggregation to reduce the payload size and optimize transmission efficiency, especially for environmental or energy monitoring applications.
    \end{itemize}
\end{itemize}

\subsubsection{Campus Scenario for the Application Layer}

For example, the Application Layer can support a campus-wide monitoring application that aggregates data from temperature, humidity, and air quality sensors across various campus locations~\cite{journals/csur/MishraG23,conf/wcnc/ThangadoraiSGK24,journals/tpds/MishraGBD24,journals/wpc/ChopadeGD23}. This allows real-time insights into environmental conditions for improved campus safety.

\subsubsection{Challenges and Solutions}

\begin{itemize}
    \item \textbf{Data Aggregation Efficiency:} With multiple applications on campus, data aggregation reduces network load by combining similar data before transmission.
    \item \textbf{Adaptive Application Support:} Different applications have varying requirements; thus, the Application Layer adapts data formats and transmission intervals to optimize performance across diverse use cases.
\end{itemize}

\begin{table*}[h]
\centering
\caption{Layer-wise Summary of LoRaWAN-MANET Architecture for Campus Setting}
\begin{tabular}{|p{3cm}|p{5cm}|p{7cm}|}
\hline
\textbf{Layer} & \textbf{Key Features} & \textbf{Campus Scenario and Solutions} \\
\hline
\textbf{Physical Layer} & 
\begin{itemize}
    \item Spread Spectrum Modulation (CSS)
    \item Long-range communication (several kilometers)
    \item Low power consumption
\end{itemize} & 
\begin{itemize}
    \item Campus-wide communication with minimal power usage, ideal for battery-operated devices
    \item Reduced infrastructure needs in outdoor spaces
\end{itemize} \\
\hline
\textbf{Data Link Layer} & 
\begin{itemize}
    \item Error detection and correction
    \item Aloha-based MAC protocol
    \item Duty cycle management
\end{itemize} & 
\begin{itemize}
    \item Controls medium access for reliable data transmission
    \item Mitigates collision in crowded campus environments
\end{itemize} \\
\hline
\textbf{Network Layer} & 
\begin{itemize}
    \item Routing protocols (e.g., AODV)
    \item Mobility management
\end{itemize} & 
\begin{itemize}
    \item Supports seamless mobility, ensuring connectivity as nodes move across campus
    \item Dynamic routing adapts to topology changes
\end{itemize} \\
\hline
\textbf{Transport Layer} & 
\begin{itemize}
    \item Reliable data transfer
    \item Flow control and congestion management
\end{itemize} & 
\begin{itemize}
    \item Ensures delivery of critical data (e.g., alerts) even under high network load
    \item Flow control prevents data loss in crowded areas
\end{itemize} \\
\hline
\textbf{Application Layer} & 
\begin{itemize}
    \item Multi-application support (e.g., monitoring, alert systems)
    \item Data aggregation and compression
\end{itemize} & 
\begin{itemize}
    \item Supports various campus applications, from monitoring to emergency alerts
    \item Data aggregation reduces network load, optimizing performance
\end{itemize} \\
\hline
\end{tabular}
\label{tab:LoRaWAN_MANET_layers}
\end{table*}

\section{Challenges in LoRaWAN-MANET Architecture}

The integration of LoRaWAN with Mobile Ad-hoc Networks (MANETs) for campus-wide applications brings several unique challenges. This section discusses the primary issues faced in deploying LoRaWAN-MANET architecture, categorized by various network layers, focusing on aspects such as energy efficiency, data reliability, and network scalability.

\subsection{Physical Layer Challenges}
\begin{itemize}
    \item \textbf{Signal Interference:} In a densely populated campus environment, interference from Wi-Fi networks, cellular signals, and other wireless devices can degrade LoRaWAN’s signal quality, especially in indoor spaces.
    \item \textbf{Range Limitations in Urban Layouts:} While LoRaWAN can cover several kilometers in open areas, buildings and other physical obstructions on campus can impact signal propagation, requiring additional nodes or repeaters to maintain connectivity.
    \item \textbf{Energy Constraints:} Devices in a LoRaWAN-MANET architecture often operate on limited power sources, like batteries, making energy management essential to maximize network lifespan.
\end{itemize}

\subsection{Data Link Layer Challenges}
\begin{itemize}
    \item \textbf{Medium Access Control (MAC) Collisions:} The Aloha-based MAC protocol, though simple, can result in frequent data packet collisions, especially in high-density scenarios like classrooms or events, leading to communication delays and packet loss.
    \item \textbf{Bandwidth Limitations:} LoRaWAN’s low data rate can be a constraint for applications requiring higher bandwidth, such as streaming or real-time monitoring.
    \item \textbf{Duty Cycle Restrictions:} Regulatory duty cycle limits may restrict the number of transmissions a device can make, impacting applications that require frequent updates.
\end{itemize}

\subsection{Network Layer Challenges}
\begin{itemize}
    \item \textbf{Dynamic Topology Adaptation:} Campus environments often experience frequent changes in network topology due to the mobility of students and staff, requiring adaptable routing protocols that can quickly reconfigure paths~\cite{7803607,7815384,7880946,8016573,8170296,8480255}.
    \item \textbf{Energy-Efficient Routing:} Given the limited power resources, finding routes that minimize energy consumption while maintaining reliable communication is essential.
    \item \textbf{Latency in Multi-hop Communication:} Routing data over multiple hops can introduce latency, which may impact time-sensitive applications like emergency notifications.
\end{itemize}

\subsection{Transport Layer Challenges}
\begin{itemize}
    \item \textbf{Reliable Data Transfer:} Ensuring data delivery in a highly mobile and dynamic environment can be challenging, especially for critical data like alerts. Packet loss due to congestion or disconnections can impact application performance.
    \item \textbf{Flow Control and Congestion Management:} With potentially thousands of connected devices, managing data flow to prevent congestion and packet loss is critical, particularly for applications with periodic data requirements.
\end{itemize}

\subsection{Application Layer Challenges}
\begin{itemize}
    \item \textbf{Data Aggregation Efficiency:} High volumes of data from multiple sensors and devices can lead to network congestion. Aggregating data efficiently without losing critical information is challenging.
    \item \textbf{Adaptability to Diverse Applications:} Campus applications range from low-latency alert systems to high-bandwidth monitoring, requiring the Application Layer to adapt data handling and prioritization to support different needs.
    \item \textbf{Privacy and Security Concerns:} In a campus setting, sensitive data (e.g., health or security alerts) may be transmitted over LoRaWAN-MANET, necessitating encryption and secure data handling protocols to ensure privacy.
\end{itemize}

\section{Advantages of LoRaWAN-MANET Architecture}

The integration of LoRaWAN with Mobile Ad-hoc Networks (MANETs) presents several significant advantages, especially in a campus environment. This section highlights the key benefits, focusing on aspects such as scalability, energy efficiency, flexibility, and application diversity.

\subsection{Scalability}
\begin{itemize}
    \item \textbf{Support for Numerous Devices:} LoRaWAN’s design allows for a large number of connected devices within a network, making it ideal for campus applications where thousands of students, faculty, and devices may need to communicate simultaneously.
    \item \textbf{Dynamic Network Expansion:} The infrastructure-less nature of MANETs enables easy addition of new nodes without the need for centralized control, facilitating network scalability as campus needs evolve.
\end{itemize}

\subsection{Energy Efficiency}
\begin{itemize}
    \item \textbf{Low Power Consumption:} LoRaWAN is specifically designed for low power usage, making it suitable for battery-operated devices. This reduces maintenance costs and extends the operational life of sensor nodes across campus.
    \item \textbf{Adaptive Power Management:} MANETs can implement adaptive power management strategies to optimize energy consumption based on the network conditions and application requirements, further enhancing energy efficiency.
\end{itemize}

\subsection{Flexibility and Resilience}
\begin{itemize}
    \item \textbf{Infrastructure Independence:} LoRaWAN-MANET architecture can function without a fixed infrastructure, providing robust communication in various scenarios such as emergency situations, outdoor events, or remote areas of the campus.
    \item \textbf{Dynamic Topology Management:} The ability of MANETs to self-organize and self-configure allows the network to adapt seamlessly to changes in topology, such as mobile users and varying environmental conditions, ensuring continuous communication.
\end{itemize}

\subsection{Cost-Effectiveness}
\begin{itemize}
    \item \textbf{Reduced Infrastructure Costs:} The infrastructure-less characteristic eliminates the need for expensive base stations and wiring, leading to lower deployment costs and easier installation of communication networks on campus.
    \item \textbf{Long-Term Savings:} Energy-efficient operation combined with minimal maintenance requirements results in long-term savings for campus management and resource allocation~\cite{9479778,9311219,9166711,9164991,9130098,8502812,8326735}.
\end{itemize}

\subsection{Diverse Application Support}
\begin{itemize}
    \item \textbf{Wide Range of Use Cases:} LoRaWAN-MANET architecture supports various applications such as environmental monitoring, smart campus services, emergency alerts, and asset tracking, catering to the diverse needs of campus life.
    \item \textbf{Real-Time Communication:} The integration facilitates real-time data transmission for critical applications, such as safety alerts or system monitoring, enhancing campus safety and operational efficiency.
\end{itemize}

\subsection{Enhanced Security Features}
\begin{itemize}
    \item \textbf{Data Encryption:} LoRaWAN supports end-to-end encryption, ensuring the confidentiality and integrity of data transmitted across the network, which is vital for protecting sensitive information within campus applications.
    \item \textbf{Network Resilience to Attacks:} The decentralized nature of MANETs adds a layer of resilience against network failures or attacks, improving overall security in a campus environment.
\end{itemize}

\section{Conclusion}

The integration of LoRaWAN with Mobile Ad-hoc Networks (MANETs) offers significant potential for enhancing communication networks within campus environments. This study has demonstrated that the combined architecture leverages the strengths of both technologies, providing a scalable, energy-efficient, and flexible solution that can address the diverse communication needs of educational institutions. The advantages of this integration include the capability to support numerous connected devices, reduced infrastructure costs, and improved real-time communication for critical applications such as emergency alerts and environmental monitoring. However, challenges such as signal interference, data packet collisions, and energy management must be addressed to ensure the reliability and performance of the network. Future research should focus on developing innovative routing protocols, adaptive energy management strategies, and robust security measures tailored to the dynamic nature of campus environments. By addressing these challenges, the LoRaWAN-MANET architecture can be further optimized, paving the way for the deployment of advanced communication systems that enhance campus life and safety. In conclusion, this paper underscores the transformative potential of LoRaWAN-MANET integration in creating resilient, efficient, and cost-effective communication networks, ultimately contributing to a more connected and responsive campus experience for all users.

\bibliographystyle{IEEEtran}
\bibliography{Paper.bib}

\end{document}